\documentclass[traditabstract]{aa}
\usepackage{graphicx}
\usepackage{txfonts}
\begin{document}
   \title{A nova origin of the gas cloud at the Galactic Center ?}

   \author{F. Meyer and E. Meyer-Hofmeister}

   \institute
     {Max-Planck-Institut f\"ur Astrophysik, Karl-
     Schwarzschild-Str.~1, D-85740 Garching, Germany}
\date{}

\abstract
{The recent discovery by Gillessen and collaborators of a cloud of gas falling 
 towards the Galactic Center on a highly eccentric orbit, diving
 nearly straight into the immediate neighborhood of the central
 supermassive black hole, raises the important question of its
 origin. Several models have already been proposed. 
 Here we suggest that a recent nova outburst has ejected a ring-like 
 shell of gas. Viewed at high inclination, that could account
 for the mass, head and tail structure, and the unusually high eccentricity
 of the observed cloud in a natural way, even as the nova moves
 on an orbit quite normal for the young stars in the close
 neighborhood of the Galactic Center. We illustrate this by
 calculating orbits for the head and tail parts of the ejecta and the
 nova that has produced it. We briefly discuss some of the questions
 that this model, if true, raises about the stellar environment close to
 the Galactic Center.
}
 

\titlerunning {A nova origin of the gas cloud at the Galactic Center?}
\maketitle
%
\section{Introduction}
The discovery  by Gillessen et al. (2012) of a cloud of gas freely
falling toward the Galactic
center on an orbit diving deeply into the immediate neighborhood of the
central supermassive black hole, is a great
achievement that raises important questions for our understanding of
the Galactic Center. The cloud was recognized as a moving object
in L'-band observations. In the position-velocity maps of Br$\gamma$ 
emission, obtained with SINFONI, a bright `head' of emission appears 
together with a `tail' of lower surface brightness (Gillessen et
al. 2012, Fig.2). How did this cloud form and how did it get the peculiar orbit so close to the central supermassive black hole
that it already now shows the destructive shear of gravity? The surprising
answer might be: We are witnessing the recent outburst of a nova, the
ejecta now freely falling toward the Galactic Center and being
observed as they disperse in the close neighborhood of the black hole.

\section{Proposed models}
The origin of this cloud is an intriguing question, and several
proposals immediately have been made.
Gillessen et al. (2012) present two suggestions. 
The cloud could originate from (1) colliding winds
in the so-called `clock-wise' disk of young, massive 
stars, which might create low angular momentum gas, which then falls into the 
potential well of the black hole, or (2) from a compact planetary nebula.
The work of Burkert et al. (2012), which
includes physical processes in the cloud/atmosphere system, 
focuses on detailed numerical simulations to follow the cloud on its
way towards its pericenter and beyond. Additional hydrodynamical
simulations were carried out by Schartmann et al. (2012) to compare
the validity of the two suggested models, the
{\it{Compact Cloud}}  \rm{scenario and the } \it{{Spherical Shell}}
\rm {scenario}. The latter model is favored since their test
particle computations of an initial ring structure at the apocenter
(in their two-dimensional simulation) allow parameters to be
chosen such that the hydrodynamical evolution for the following
years, showing the severe stretching of the ring by the gravitational
force of the black hole, matches the observed structure of the cloud in the
position-velocity diagram.

An alternative third picture has been proposed by Murray-Clay \&
Loeb (2012): a dense, proto-planetary disk around a star,
scattered from the observed ring of young stars, which loses mass through
photo-evaporation as it approaches the supermassive black
hole. Taking up the 
photo-evaporation suggestion, Miralda-Escud\'e
(2012)  presents a similar model, proposing that this process happens in
an old, low-mass star,
deflected into its high-eccentricity orbit after a close
encounter with a stellar mass black hole. The small debris disk around the
star could create a cloud like the one observed at every orbit around the
Galactic Center. The latter paper includes a detailed discussion of the
problems faced by the various proposed models.

\section{ Mass of the cloud, velocity and the warm dust}

There are three features that may point to another origin, a nova 
outburst: a mass of the cloud of $10^{-5}$ solar masses, dust, and 
velocities on the order of 1000 km/s. 

Nova outbursts are a much studied phenomenon of
cataclysmic variables. These close binaries contain a white dwarf
primary and usually a low-mass Roche-lobe-filling secondary star, from which 
mass flows over to the primary star via an accretion disk. If enough
matter has accumulated on the surface of the white dwarf, a thermonuclear
run-away explosion in the partially degenerate gas, the nova outburst,
occurs. These outbursts occur again and
again over the lifetime of the binary system.

Observed nova shells have masses between $10^{-5}$ and $10^{-4}$ solar
masses (Shore 2008) and ejecta velocities of a few hundred up to more
than a
thousand km/s (Downes \& Duerbeck 2000). The observed cloud mass of about $10^{-5}$ solar masses
is at the lower limit of shell masses and thus might be only a
fraction of the total ejected mass. The three-dimensional velocity of the cloud
was 1200 km/s observed in 2004 and increased to 2350 km/s in 2011 as
it moved closer to the central black hole (Gillessen et al. 2012). 
The observed velocities would result from the addition of orbital and ejecta
velocity of the nova. That these are found to be of the same order of
magnitude is an interesting fact as we see later.

Nova shells contain dust in very different amounts. 
The dust formation theory is complex, since for molecules to form and
grains to aggregate, temperatures below 1000K are required. The process 
depends on the metal abundances, and these vary between the lower mass CO white
dwarfs and the more massive ONe white dwarfs.
Gillessen et al. (2012) estimate the temperature and the amount of
warm dust in the cloud from the L'-band (3.76$\mu$m) luminosity as the 
reradiation by dust grains, irradiated by the massive stars in the
surroundings, as 550$\pm$ 90 K and $2\times 10^{23}$ g. The latter is
on the lower side
of the observed range in nova shells. We note that destruction of dust in the
evolution of nova shells has been discussed (Evans et al. 2005)

\section{Geometry of the expanding nova shell}

Observations of resolved nova shells document a variety
of shapes of ejected gas around the white dwarf (O'Brian \& Bode
2008). Slavin et
al. (1995) analyzed shells of 13 classical novae and
found indications of a correlation between the speed class of a novae 
(which measures the rate of luminosity increase in outburst) and the shape
of the shell, in the way that faster novae tend to comprise randomly
distributed clumps of ejecta superposed on spherically symmetric
diffuse material, while slower novae produce more structured
ellipsoidal remnants with rings of
enhanced emission. Downes \& Duerbeck (2000) studied images of
resolved shells of recent novae, and find the densest gas clouds
often concentrated in a ring. Krautter et al. (2002) obtained
near-infrared images of classical novae, and find further support
for a correlation between speed class and shape of the shell, and
also point out a more or less pronounced density inhomogeneity.

As an example, we consider here a ring-like shape of the ejecta.
The question arises how the orbital plane of the binary system 
(the preferred direction of the ejected ring) is oriented with
respect to the observer. Depending on the inclination, our
line of sight passes through a higher column density at the `edge' of
the ring (i.e. at the ends of the major axis of the ring projection). 
A simple slab model with thickness $d$ (the diameter of the ring
cross section) and inclination $\vartheta$ yields a length $l$
of this column $l=d /\cos\vartheta$. For the stochastic mean of
arbitrary orientation, $\vartheta=60^\circ$, $l/d$ is 2.0, however it 
rises rapidly
with higher inclination, $l/d=5.8$ for $\vartheta=80^\circ$. For
$\vartheta=90^\circ$, a ring seen edge-on, for a diameter 1/10 of
the ring inner radius, $l/d$ has a maximum value of 9.2. Because the surface
brightness of the
ring emission is proportional to the column density, the observations
of Gillessen et al. (2012), which show a pronounced head-tail structure
in brightness distribution, would point to a rather high inclination.
We consider a ring in the orbital 
plane of the observed
cloud, which {\it{is}} \rm{highly inclined to our line of sight. 
We note that there is no reason why the two planes should coincide, so
it is a somewhat arbitrary choice made for the sake of simplicity.

In their {\it{Spherical Shell}} \rm{scenario SS0 for the origin of 
the cloud Burkert et al. (2012) and Schartmann et al. (2012)
also considered 
a gas shell starting at the apocenter of the observed highly
eccentric orbit of the cloud. In their two-dimensional simulation they
followed the evolution of an initial ring of test particles in the
orbital plane. The distortion of this ring into an elongated structure
can be seen in their Figs. 12 and 5, respectively. The initial
conditions, an expansion velocity of 125 km/s, radius, and 
thickness of the ring, were chosen to best reproduce the observations
of the cloud.

\begin{figure*}
   \centering
   \includegraphics[width=12cm]{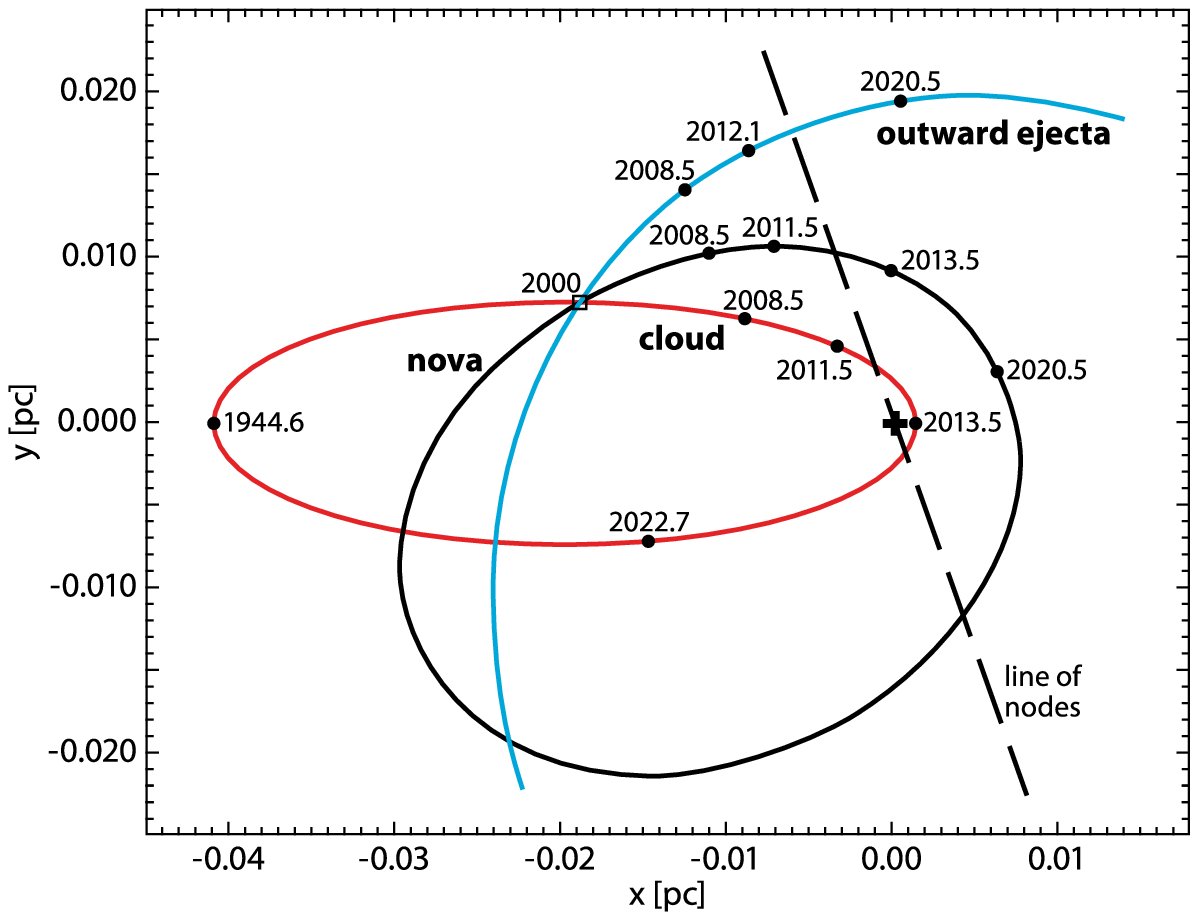}
     \caption{ 
       Three orbits around the supermassive black hole at the Galactic
       Center. Red line:
       Orbit of the cloud as determined in Gillessen et al.(2012)
       (taken from Schartmann et al. 2012, G2 in their Fig.1), in our model
       interpreted as the inward ejecta of highest surface brightness
       for the observer. Black: the
       orbit of the nova that produced it. Blue:
       the orbit of the outward ejecta, which in projection are the
       highest surface brightness feature of the opposite ring region, 
       identified with the tail. The years 2013.5, 2020.5,
       and 2012.1 denote the pericenters of the ellipses with quite
       different orientation and eccentricity, the cross denotes the
       position of the black hole. Dashed line: line of nodes, the
       intersection of the common plane of all orbits and the plane of
       sky, inclined to each other by a nearly right angle of
       i=109.6$^\circ$. See text.
       }       
       \label{f:cloud}

   \end{figure*}

\section {Three elliptical orbits around the black hole}

In Fig. \ref{f:cloud} we show three orbits. One is the orbit of the
cloud as derived by Gillessen et al. (2012) from the observations. 
There are many ways in which this cloud orbit can be thought of as having
come about as the best visible part of nova ejecta. The two other
orbits are such an example, and they show the orbit of the nova system
itself and a corresponding, in projection to the observer,
`second-best' visible part of the ejecta.

\subsection{An example}

We note that there is a wide range of parameters in shell mass, ejecta
velocity, shape and clumpiness of observed novae. Besides the
orientation of the plane in which the binary lies, another free
parameter for our modeling is the moment of the nova explosion. 

The explosion should have happened before
the cloud was first observed in the year 2002.
A nova outburst right in 2001 seems excluded,
since during the nebular phase the hydrogen burning, the white
dwarf surface is exposed and irradiates the expanding dust shell, and
the re-radiated infrared luminosity would significantly decrease when the 
distance between central star and shell grows as the
shell expands, which does not seem to be observed. Thus one
concludes that the hydrogen surface burning, typically continuing for 
two to three years after the onset of the outburst, had already ended and
the outburst should have occurred at or before the year 2000. (The same date
was chosen for the start of the evolution of a test particle cloud 
in the analysis of Gillessen et al. 2012.) 

With our assumption of the ring lying in the orbital plane of the observed
cloud, we have to find the orbit of the nova in this plane such
that the orbit of the ejecta, best visible in projection, is that of 
the cloud. We take the ejecta velocity as 500 km/s corresponding
to a slow nova. This velocity is the difference between the
velocities of the cloud and the nova and, for the best visibility in
projection, is directed at a right angle to the line of sight, as
discussed in the previous section. The same procedure also yields
the orbit of the second part of the ejecta at which we again look
tangentially through the edge of the ring, directed outward with
respect to the orbit around the black hole. Figure\ref{f:cloud} shows
the resulting orbits: The red line
is the orbit of the inward directed ejecta, identified with the
observationally determined orbit of the cloud, black is the orbit of
the nova from which the ejecta originated, and blue is the orbit of
the outward directed ejecta.

\subsection{Orientation and eccentricity of the orbits}

 Table 1 lists the parameters of these three orbits. The pericenter of
the ellipses are reached for the cloud in 2013.5 and the nova in 
2020.5 and was already reached for the outward ejecta in 2012.1. 
The major axis of the binary and
the outward ejecta are turned by 26.0$^\circ$ and -117.8$^\circ$,
respectively, compared to the cloud. Since the orbits of the nova and
the outer ejecta lie in the same plane as the orbit of the cloud, their
inclination i and their position angle of ascending node $\Omega$ are
the same, i=106.55$^\circ$ 
and $\Omega$=101.5$^\circ$. 

But the most interesting point is the very different eccentricity.
In contrast to the extreme value of the cloud, e=0.9384, the values 0.628
and 0.459 are quite normal for stellar orbits in the neighborhood of
the Galactic Center, as compiled by Gillessen et al. (2009). Our example shows 
that if the cloud originates in a nova outburst, its high
eccentricity can result in a natural way. Generally, ejecta in
the direction towards the 
inside of an orbit will move on orbits with higher eccentricity, and
ejecta in direction outwards with lower eccentricity.

\begin{table}
\caption{\label{t:par} \bf{Keplerian orbits around the black hole}}
\begin{center}
\begin{tabular}{|l|l|l|l|}
\hline
&&&\\

\  parameters & inward ejecta & \,\,\,\,\,\,binary & outward ejecta\\
&= cloud &&\\
\hline
&&& \\
 semi-major axis a &$6.5\times10^{16}$cm & 6.0$\times 10^{16}$cm & 
                                     1.1$\times10^{17}$cm\\
 eccentricity e & 0.9384&0.628&0.459\\
 orbital period &137.9 years &122.8 years &300.4 years\\
 time of pericenter &2013.5&2020.5&2012.1\\
 pericenter distance & & &  \\
 \,\, from black hole &4.0$\times 10^{15}$cm &2.2$\times 10^{16}$cm&
                                    5.9.$\times 10^{16}$cm\\
&&& \\ 
\hline 
\end{tabular}
\end{center}

References: Values for the cloud from Gillessen et al. (2012), values
for the binary and outward ejecta this paper.
\end{table}

\section{Br$\gamma$ radiation} 
In the observations of Gillessen et al. (2012), the very prominent
cloud head is visible in both the infrared
L'-band continuum and the Br$\gamma$ line, and the lower surface brightness
feature, the tail, is still clearly visible in Br$\gamma$
radiation. Its integrated  Br$\gamma$ flux is comparable to that of
the cloud. Weak Br$\gamma$ light distributed between the two can be
seen. We have not calculated
the orbits of other ejecta in our ring-like shell model, but do
tentatively identify this with light from the other ring regions
between the two end points. One would expect a certain broadness in
the line of sight velocity distribution and, if one wishes, one might
see such a feature in the position-velocity maps (Fig. 2 of Gillessen
et al. 2012), in particular that
of 2011. In addition there appears to be some indication of clumpiness.

While the head between 2008 and 2011 had moved
significantly closer to the Galactic Center and becomes distorted
along its orbit by the tidal forces, the tail feature in projection has
not changed so much in position and velocity.
We compare this development of the cloud structure with the orbits shown in our
Fig.\ref{f:cloud}. Since we look closely along the
common orbital plane, the projections occur practically parallel in the
direction vertical to the line of nodes. The change in positions on
the orbits projected on the plane of the sky (perpendicular to the
line of sight) and the velocities projected along the line of sight
on the two orbits, named cloud and outer ejecta, show such 
behavior. 

If we suppose a common origin of both the head and tail of the cloud, 
what is the velocity with which those two parts 
moved apart from each other? A look at Fig.\ref{f:cloud} shows that
the true distance between the two in 2008 is very close to that
projected on the line of nodes, i.e. very close to the one projected on the plane
of the sky. From Fig. 2 of Gillessen et al. (2012) the latter is about
0.25 arcsec, corresponding to $10^{16.5}$ cm at the distance of the
Galactic Center. For the eight years between the assumed time of explosion in
2000 and the observation in 2008, this yields a velocity of about 1200 km/s for
the speed of separation, or an ejecta velocity of 600 km/s. This is 
a typical value for observed nova ejecta and is indeed close to the
velocity assumed in our example. A similar value results for the 2011 data,
maybe slightly affected by the gravitational acceleration
from the Galactic Center.

In an expanding nova shell the density decreases 
with time, which leads to a gradual decrease in brightness in the
recombination lines. This affects the outer ejecta that in our
example find themselves on a wide orbit around the black hole. In contrast
the inner ejecta, the cloud head, fall
on a highly elliptical orbit nearly straight towards the black
hole. This part of the ejecta becomes elongated along its path by the
tidal force, but becomes compressed in the two orthogonal directions
so that the density even increases. The general picture of the cloud
region is that of a HII region of roughly constant temperature, which
is kept ionized by the radiation field of the surrounding bright
stars (Gillessen et al. 2012). In this situation the recombination
rate, hence the emission of Br$\gamma$ line radiation, is proportional
to the square of the density. This might account for the contrast in
surface brightness of the Br$\gamma$ radiation as a result of the
different densities in head and tail. We notice that large-scale
clumpiness of the ejecta might also affect the appearance of the
general cloud pattern.

Finally, the question arises whether the nova itself could be seen. After
the years since about 2000 the nova is probably extinct. To now
observe Br$\gamma$ radiation from the nova would require a
remaining, significantly less massive, but denser HII region surrounding 
the binary system that is moving together with it.

\section{Cataclysmic variables near the Galactic Center?}

If true, the nova model, demanding the presence of a cataclysmic
variable so close to the Galactic Center, raises interesting
questions. Formation of a CV white dwarf needs
about $10^8$ years of evolution into the asymptotic giant branch phase 
of a four to five solar mass main sequence star (Ekstr\"om et al. 2012). 
Furthermore, a low-mass star of mass less than that of the white dwarf must
be captured to form a cataclysmic
variable, which then could live for up to Gyrs, depending on the white
dwarf mass and the mass transfer rate from the secondary star. At present
there seems to be a dearth of low-mass red giants in the Galactic
Center. It has been suggested that encounters in the dense stellar
environment may remove the giant envelope, and thus reduce the number
of observable low-mass giants (Lacy et al. 1982, Genzel et al. 2010). 

Two processes are 
commonly thought to lead to the formation of a close binary system
with a white dwarf primary: (1) standard common envelope evolution of
an originally wide binary during the giant phase of the primary star
(see e.g. Nelemans \& Tout 2005), and (2), capture of a low-mass star 
in dense stellar environments, by close encounters with the white dwarf.
In the bulge of M31 the latter process, dynamical formation, is well
documented for the overabundance of low-mass X-ray binaries (Voss \&
Gilfanov 2007), with a neutron star or stellar
mass black hole primary instead of the white dwarf. Similarly, in the
Galactic Center Muno et al. (2005) found a 20-fold excess of X-ray
transients in the central parsec with respect to the larger 10 parsec 
environment.

A particular channel in such environments, open only to white dwarf
binary formation, is the tidal capture of a low-mass star by the giant
envelope of an evolved star that already has the degenerate core of white
dwarf size at its center. The ensuing common envelope phase will form
a pre-CV or CV type close binary, and, as discussed above, at the same
time remove the original red giant from view. 

Such processes could then
lead to a significant overabundance of nova systems in the very central
cusp of our galaxy.

\section{Conclusions}

We have investigated the possibility that the cloud observed by
Gillessen et al. (2012) in the close neighborhood of the Galactic
Center results from a nova outburst. Mass, velocities, and the presence
of dust agree with those of observed nova shells. We find that 
simple examples for the orbits of ejected matter allow 
interpretation of the cloud head in its observed peculiar orbit around the
Galactic Center as part of the expanding shell of a nova that
exploded around the year 2000. We note that different parts of a ring-like shell
will appear at different brightness to the observer, the cloud head being the
brightest one. The velocity with which the observed
parts of the cloud structure, head and tail, move
apart from each other, suggests that the tail is the, for the observer, 
second brightest part of the shell, ejected in the direction opposite
to that of the head.
The high eccentricity of the cloud head can be understood as caused by
the addition of the velocities of the nova itself and the ejecta in
that particular direction.

Though the nova origin model seems promising it still needs
further evaluation. 
We note a wide range of possible parameters for ejected nova shells and
their orientation with respect to the observer, which could allow a
detailed comparison of such a model with the observations of that
extraordinary cloud event.

\end{document}